\documentclass{ws-ijgmmp}

\begin{document}

\markboth{Aneta Wojnar}

%
\catchline{}{}{}{}{}
%

\title{White dwarf stars in modified gravity
}

\author{Aneta Wojnar}

\address{Laboratory of Theoretical Physics, Institute of Physics, University of Tartu,
W. Ostwaldi 1, 50411 Tartu, Estonia\\
\email{aneta.magdalena.wojnar@ut.ee}}

\maketitle

\begin{history}
\received{(Day Month Year)}
\revised{(Day Month Year)}
\end{history}

\begin{abstract}
A few questions related to white dwarfs' physics is posed. It seems that the modified gravity framework can be a good starting point to provide alternative explanations to cooling processes, their age determination, and Chandrasekhar mass limits. Moreover, we have also obtained the Chandrasekhar limit coming from Palatini $f(\mathcal{R})$ gravity provided by a simple Lane-Emden model.
\end{abstract}

\keywords{stellar structure; white dwarfs; modified gravity.}

\section{Introduction}
According to the current most reliable knowledge, the final evolutionary state of main-sequence stars leads
to ultra-dense stellar remnants which are commonly referred to as compact objects. Depending on the star's
initial mass, the stellar object can evolve under a self-gravitational collapse process into a white dwarf, a
neutron star, or a black hole. As black holes provide no direct electromagnetic radiation (besides their
shadows \cite{ab1,aki}), they are rather difficult objects to be used in order to test and compare gravitational theories,
where only gravitational waves provide reliable information. On the contrary, neutron stars, can be observed
not only via gravitational waves \cite{ab2} but also through electromagnetic radiation \cite{hew1,hew2}, allowing to constrain models
of the nuclear matter at ultra-high densities as well as our models of the gravitational interaction. A typical
neutron star has a mass of $1.5$ solar masses, and a radius around $10$ kilometers, making it the smallest and
densest stars known so far. Their structural properties can be roughly understood by assuming that they are
composed of neutrons supported by their degeneracy pressure (Pauli exclusion principle), which makes them
particularly suitable to test our theories of gravity and matter at the highest densities achievable in the
universe. Indeed, the actual matter composition at densities above the saturation density $\sim2.8\times10^{17}\frac{\text{kg}}{\text{m}^3}$ at
the star's center is still unknown, which is a crucial element to modeling such objects. The recent observation
of gravitational waves and electromagnetic radiation of a neutron star merger, and its ability to rule out
certain families of modified gravity theories \cite{ez,bor,san,ab3,jana} is a clear example of their relevance for current research.
Despite this success, there is still a long way to go in order to extract useful information about the internal
structure, dynamics and composition of neutron stars. This fact introduces a fundamental degeneracy, since
we cannot distinguish whether the differences in the mass-radius profiles are caused either by the equation of
state assumed or from the theory of gravity from which the hydrostatic equilibrium equation was obtained.

White dwarf stars \cite{shap,koe} are also the final state of the stellar evolution but for main sequence stars whose masses
lie below $8-10$ solar masses. When the stage of hydrogen-fusing of a main-sequence stars ends, it leads into
an expansion to a red giant star. Subsequently, helium is fused to carbon and oxygen, which are building up
at the star's center. However, the core temperature is not enough to further fuse carbon (the temperature
depends on the red giant's mass) and thus the outer layers, composed of lighter elements, will be ejected
under the form of a planetary nebula, leaving behind a carbon-oxygen core. Since there is no further fusion
reaction in a new-born white dwarf providing energy to support the star against gravitational collapse, the
star is supported by electron degeneracy pressure only. After the planetary nebula is formed, the mass of a
white dwarf becomes around $1$ solar mass packed into an Earth-size volume, making up densities around $10^9 \frac{\text{kg}}{\text{m}^3}$ , and including it into the family of compact stars. The same intense radiation that makes the planetary
nebula form will cool down the white dwarf causing the core material to eventually crystallize. The cooling
process of such an object is very long. In order to become a cold black dwarf (a star which does not emit any
heat or light and therefore would be undetectable by optical observations), the white dwarf would need more
time than the current age of the Universe. All these elements make white dwarfs important objects to study
since the cooling process provides information on the star's formation in the Milky Way.

There are two main problems related to the physics of white dwarfs: their age (related to the very low-mass) and super-Chandrasekhar masses, which are hard to explain by current astrophysical models. It indicates that we are far from understanding these objects; it may also mean that there are processes in the stellar evolution which we do not take into account.

The very curious case is a discovery of a
very low mass white dwarf with 0.2 solar masses in the binary system KIC 8145411 \cite{masuda}. According to the
commonly accepted models \cite{la}, such a white dwarf would have to evolve from a very small progenitor
star which had had to be older than the Universe ($13.6$ billion years) in order to run out of its fuel and
then to follow the known scenario of turning into a white dwarf. 
The age of an observed white dwarf is roughly speaking the
sum of the lifetime of the main sequence progenitor star and of the white dwarf cooling time. 
The lifetime of the main sequence
star is proportional to $\tau\sim M^{-3}$ , where $M$ is a main sequence star's mass. Thus, the less massive the
star, the longer time needed to turn it into a white dwarf. Usually such low-mass white dwarfs are
observed in binary systems, in which their companions steal the outer layers, speeding the process of
the exposition of the core. But in the considered binary KIC 8145411 the orbit of the second
companion, being a Sun-like star, is far wider than the required one to produce such a low-mass white
dwarf via their interactions. There could have existed a third star which had changed the orbits of the
binary, making the cannibalism process possible, when the white dwarf progenitor was a red giant -
but then the current orbit would have not been nearly circular as observed. Another explanation of its
low mass could be the existence of large planets in the past which could have siphoned the gas from
the star's surface. However, none of those scenarios is satisfactory, especially in the light of other
single low-mass white dwarfs whose companions have not been yet detected, assuming that they exist.

Another ambiguity related to the age of these compact objects are discoveries of white dwarfs being older than their host cluster, assuming that the cluster's stars had come into being at the same time.
This way, there are white dwarfs such as for example WD
0346+246 \cite{ham}, SDSS J091709.55+463821.8 \cite{kilic} (which is a binary system of two low-mass white dwarfs \cite{kilic2})
that are very old - around 11-12 billion years. Usually they are very faint (as in the case of one of the
members of J091709.55+463821.8, which was not detected optically neither via X-ray and radiopulsar observations \cite{agu}, which is the reason why the companion should be a white dwarf) and cold, with
the effective temperatures very close to the limit $T=3000$K. This is a minimal temperature that this
type of star can have in order not to be older than the Milky Way, assuming that those stars are already
crystallized. It is accepted that the crystallization has to happen during the cooling period of a white dwarf - this is
a process which appears as the fluid in the star cools and as a result, turns into a solid state. However,
when fluid crystallizes it releases latent heat which must be also added in the total energy supply,
being further radiated. This increases the cooling time. Although it was shown that the crystallization 
phenomena can shorten the cooling time \cite{mestel}, this process is still a source of uncertainty in the age of cold
white dwarfs.

The lifetime of a main sequence star as well as the cooling time of white dwarfs (taking into account
the crystallization process) are quantities which depend on the luminosity and mass of the stars \cite{her}. It
was shown \cite{sak1,sak2,gonzalo,aneta4,aneta5,davis,benito} that in the case of some gravitational theories both mass and luminosity
change because they are sensitive to modifications of Einstein's gravity. Indeed, although white
dwarfs can be treated by the Newtonian hydrostatic equilibrium equation, which is used in the
computations of the lifetime of the main sequence stars and cooling processes, many theories of
gravity modify that equation (see e.g. \cite{reva} and references therein). 
Moreover, it was also demonstrated that modified gravity has also an impact on the early stars' evolution, such us for example Hayashi tracks and radiative core development \cite{aneta4,chow,chang}, which clearly indicates that re-examination of the evolutionary processes in the framework of modified gravity can provide answers to the white dwarfs' age problem.

Another important comment related to the star's age is that lithium abundance in stellar atmosphere depends on gravity model, too, introducing an additional uncertainty to age determination techniques of young stars and globular clusters since some of them rely on the lithium depletion. It was shown that in the case of Palatini $f(\mathcal{R})$ gravity the age of an individual star is significantly shorter than the age in the model provided by GR \cite{aneta5}. Therefore, when the total star's evolution is re-analyzed, it will very probably happen that the age of the KIC 8145411 white dwarf lies in the reasonable range. 

The intriguing issue also associated to the physics of white dwarfs is super luminous SNIas, e.g. \cite{howell,scalzo,hicken,yaman,tau,silv}. This suggests that the star progenitors of such events could be super-Chandrasekhar white dwarfs which exceeded the Chandrasekhar mass limit (in the Chandrasekhar's model, a white dwarf's mass cannot be more than $1.44M_\odot$ \cite{chandra}). A white dwarf, whose mass crosses this limit by siphoning a nearby star's mass, turns into a type Ia supernovea explosion. Even more curious, including the relativistic effects to the stellar equations turns out to reduce the maximum mass white dwarfs with magnetic fields \cite{bera,deh}, thus other explanations, as for example provided by modified gravity, are necessary.  

White dwarfs have already been studied in modified gravity models. In \cite{das1,das2} the authors were 
studying the $f(R)$ gravity effects on limiting mass (super-Chandrasekhar and sub-Chandrasekhar unified by the model) and its not-uniqueness, while in \cite{carv} the $f(R,T)$ model provided the WD macroscopic properties and well as lower energy density in comparison to other models. White dwarfs are also used to constrain theories of gravity, such as for example in \cite{gar1,nn1,alt,gar,cor,ben,biesiada,ben2,is,jain,sal,babi,kalita,cris,nn2,biesiada2,eslam}. Re-examined cooling and heating processes in modified gravity and dark matter models were studied for instance in
\cite{is2,pano,hor}.

In this work we will briefly introduce Palatini $f(\mathcal{R})$ gravity with the stellar equations needed to study the Chandrasekhar's model of white dwarfs. We  will solve modified Lane-Emden equation for the polytropic parameter $n=3$ in order to discuss the Chandrasekhar mass in this framework.

We use the $(-+++)$ metric signature while $\kappa^2=\frac{8\pi G}{c^4}$.

\section{Palatini $f(\mathcal{R})$ gravity}

Palatini $f(\mathcal{R})$ gravity is one of the simplest generalization of the General Relativity: the last one can be also considered in Palatini approach. It means that the assumption on the metricity of the connection is waived and therefore one deals with two independent objects, metric $g$ and connection $\hat\Gamma$. However, in the case of the linear functional $f(\mathcal{R})$,
the field equations taken with respect to the independent connection $\hat\Gamma$ provide that  the connection is the Levi-Civita 
connection of the metric $g$. Considering more general Lagrangians this is not true anymore: as a result of the field equations the connection turns out to be a Levi-Civita connection of 
a certain metric which happens to be conformally related to the metric $g$. 

To demonstrate it, we start with the action
\begin{equation}
S=S_{\text{g}}+S_{\text{m}}=\frac{1}{2\kappa^2}\int \sqrt{-g}f(\mathcal{R}) d^4 x+S_{\text{m}}[g_{\mu\nu},\psi_m],\label{action}
\end{equation}
where the curvature scalar defined as $\mathcal{R}=\mathcal{R}^{\mu\nu}g_{\mu\nu}$ is constructed with the metric $g$ and Ricci tensor $\mathcal{R}_{\mu\nu}$ built of the independent 
connection $\hat\Gamma$. The variation of (\ref{action}) with respect to the metric $g_{\mu\nu}$ gives
\begin{equation}
f'(\mathcal{R})\mathcal{R}_{\mu\nu}-\frac{1}{2}f(\mathcal{R})g_{\mu\nu}=\kappa^2 T_{\mu\nu},\label{structural}
\end{equation}
where $T_{\mu\nu}$ denotes the energy 
momentum tensor of the matter field, defined as usually by $T_{\mu\nu}=-\frac{2}{\sqrt{-g}}\frac{\delta S_m}{\delta g_{\mu\nu}}$. For now, the primes denote derivatives with respect to the 
function's argument, that is, $f'(\mathcal{R})=\frac{df(\mathcal{R})}{d\mathcal{R}}$.

On the other hand, the variation with respect to the independent connection $\hat\Gamma$ provides
\begin{equation}
\hat{\nabla}_\beta(\sqrt{-g}f'(\mathcal{R})g^{\mu\nu})=0.\label{con}
\end{equation}
From above we immediately notice that $\hat{\nabla}_\beta$ is the covariant derivative calculated with respect to $\hat\Gamma$, that is, it is the Levi-Civita connection 
of the conformal metric
\begin{equation}\label{met}
h_{\mu\nu}=f'(\mathcal{R})g_{\mu\nu}.
\end{equation}
The structural equation is obtained from the trace of (\ref{structural}) taken with respect
to the metric $g_{\mu\nu}$
\begin{equation}
f'(\mathcal{R})\mathcal{R}-2 f(\mathcal{R})=\kappa^2 T,\label{struc}
\end{equation}
where $T=g_{\mu\nu}T^{\mu\nu}$ is the trace of the energy-momentum tensor. For particular functionals $f(\mathcal R)$ one may find a solution of the structural equation (\ref{struc}) in the form of the relation $\mathcal{R}=\mathcal{R}(T)$, allowing to express 
$f(\mathcal{R})$ as a function
of the trace of the energy momentum tensor $T$.

 One may also express the fields equations as dynamical equations for the conformal 
 metric $ h_{\mu\nu}$ \cite{DeFelice:2010aj,BSS,SSB} and the scalar 
 field, which is algebraically related to the trace of the energy momentum tensor,
 defined as $\Phi=f'(\mathcal{R})$:
 \begin{subequations}
	\begin{align}
	\label{EOM_P1}
	 \bar R_{\mu\nu} - \frac{1}{2} h_{\mu\nu} \bar R  &  =\kappa^2 \bar T_{\mu\nu}-{1\over 2} h_{\mu\nu} \bar U(\Phi)
	\end{align}
	\begin{align}
	\label{EOM_scalar_field_P1}
	  \Phi\bar R &  -  (\Phi^2\,\bar U^(\Phi))^\prime =0
	\end{align}
\end{subequations}
where the potential $\bar U(\Phi)=\frac{\mathcal{R}\Phi-f(\mathcal{R})}{\Phi^2}$ was introduced together with an appropriate energy momentum 
tensor $\bar T_{\mu\nu}=\Phi^{-1}T_{\mu\nu}$. This approach to the Palatini gravity significantly simplify studies on some particular physical problems, as shown in, e.g.
 \cite{menchon, afonso3, afonso4, sporea, aneta, aneta2, artur, gonzalo, afonso5}.

 It is also worth to comment that in the vacuum the Palatini gravity turns out to be Einstein vacuum solution with the cosmological constant. It does not depend on the $f(\mathcal{R})$ form - it can be easily demonstrated from the structural equation (\ref{struc}). Besides, in the case of analytic functionals $f(\mathcal{R})$ the center-of-mass orbits are the same as in GR \cite{junior} while the modifications of energy and momentum which appear in Euler equation are not sensitive to the experiments performed for the solar system orbits. That can change when atomic level experiments are 
available \cite{sch,ol1,ol2}.

Let us stress that Palatini $f(\mathcal{R})$ gravity differes significantly from the metric approach in which the field equations result as the $4$th order differencial equations for the metric tensor, providing different cosmological and astrophysical scenarios (see eg. \cite{cap0}). However, the extra degree of freedom related to the curvature scalar is present when the equations are transformed to the scalar-tensor representation of the metric theory, providing the 2nd order differential equations for the metric tensor. It clearly demonstrates the difference between the metric and Palatini approach since in the last one there is no extra degree of freedom.

\subsection{Non-relativistic stars for quadratic Palatini gravity}
Before going to the stellar equations, let us briefly comment that
the stellar structure in the metric-affine theory (for detailed review see \cite{reva}) was studied mainly for spherical-symmetric solutions and mass-radius
relation \cite{mr1,mr2,mr3, mr4, mr5,mr7,mr8,mr9,mr10},
the last one provided by the modified Tolman-Oppenheimer-Volkoff equations (TOV).
Some shortcomings and their solutions were considered further in \cite{ba1,ba2,ba3,pani,ba4,ba5,gonzalo2,kim,b6,gd} while problems addressed to
stability can be viewed in \cite{aneta,stab1,stab2,stab3,stab4,stab5}. Non-relativistic stars, which are here our main concern, were discussed
in \cite{aneta2,gonzalo,artur,nn1,nn2,aneta3,aneta4,aneta5,benito}.

The relativistic hydrostatic equilibrium equations (TOV) for arbitrary Palatini $f(\mathcal{R})$ gravity are given in \cite{aneta} in the form
\begin{eqnarray}
  \frac{d}{d\tilde r}\left(\frac{\Pi}{\Phi({\tilde r})^2}\right)&=&-\frac{GA\mathcal{M}}{\tilde r^2}\left(\frac{Q+\Pi}{\Phi({\tilde r})^2}\right)
  \left(1+\frac{4\pi \tilde r^3\frac{\Pi}{\Phi({\tilde r})^2}}{\mathcal{M}}\right)\label{tov_kon}\\
A&=&1-\frac{2G \mathcal{M}(\tilde r)}{\tilde r} \\
\mathcal{M}(\tilde r)&=& \int^{\tilde r}_0 4\pi \tilde{x}^2\frac{Q(\tilde{x})}{\Phi(\tilde{x})^2} d\tilde{x} \ ,\label{mass}
\end{eqnarray}
where the tilde indicates the quantities in conformal frame, 
and the generalized energy density $Q$ and pressure $\Pi$ are defined as
\begin{subequations}\label{defq}
 \begin{equation}
   \bar{Q}=\bar{\rho}-\frac{\bar{U}}{2\kappa^2 c^2}=\frac{\rho}{\Phi^2}-\frac{U}{2\kappa^2 c^2\Phi^2}=\frac{Q}{\Phi^2} \ ,
 \end{equation}
\begin{equation}
  \bar{\Pi}=\bar{p}+\frac{\bar{U}}{2\kappa^2}=\frac{p}{\Phi^2}+\frac{U}{2\kappa^2\Phi^2}=\frac{\Pi}{\Phi^2} \ .
\end{equation}
\end{subequations}
Let us recall that $\bar{U}$ and $\Phi$ depend on the choice of the gravitational model one is interested in. It is readily seen that, 
in the GR limit, $\Phi=1,\bar{U}=0$ and one recovers the standard TOV equations.

We will now consider the quadratic model
\begin{equation}\label{quad}
 f(\mathcal{R})=\mathcal{R}+\beta \mathcal{R}^2;
\end{equation}
then $\Phi=f'(\mathcal{R})=1+2\beta \mathcal{R}$. 
Using (\ref{structural}), one easily finds $\mathcal{R}=-\kappa^2 T$ such that $\Phi=1-2\kappa^2\beta T$.

Then,
the above equations in the Newtonian limit, that is, $p<<\rho$ together with
$4\pi r^3 p<< \mathcal{M}$ and $\frac{2G\mathcal{M}}{r}<<1$, in the Einstein frame are \cite{aneta2}
\begin{align}
 \frac{dp}{d\tilde r}&=-\frac{G\mathcal{M}\rho}{\Phi\tilde r^2} \ ,\\
\mathcal{M}&=\int_0^{\tilde r}4\pi x^2\rho(x)dx \ ,
\end{align}
where $\tilde r^2=\Phi r^2$ and $\Phi=1+2\kappa^2c^2\beta\rho$. Coming back to the Jordan frame we have
\begin{align}
 \frac{\Phi^\frac{1}{2}}{1+\frac{1}{2}r\frac{\Phi'}{\Phi}}\frac{dp}{dr}&=-\frac{G\mathcal{M}(r)\rho}{\Phi^2 r^2} \label{equi} ,\\
\frac{\Phi^\frac{1}{2}}{1+\frac{1}{2}r\frac{\Phi'}{\Phi}}\frac{d\mathcal{M}(r)}{dr}&=4\pi \Phi r^2\rho(r) \ 
\end{align}
which are the modified Newtonian hydrostatic equilibrium equations used to described the Newtonian stars.

\subsection{Lane-Emden equation for quadratic Palatini gravity and Chandrasekhar mass}
We will now obtain the Chandrasekhar mass as a function of solutions of the modified Lane-Emden equation. The Lane-Emden equation given in
\cite{aneta2} was obtained from (\ref{equi}) after applying the polytropic equation of state \begin{equation} 
p=K\rho^\Gamma,
\end{equation}
where $K$ and 
$\Gamma=1+\frac{1}{n}$ are polytropic constants.
 For quadratic Palatini gravity (\ref{quad}) it has the following form:
\begin{equation}\label{LEeq}
\xi^2\theta^n\Phi+\displaystyle\frac{\Phi^{-1/2}}{1+\frac12 \xi\Phi_{\xi}/\Phi}\displaystyle\frac{d}{d\xi}
\left(\displaystyle\frac{\xi^2\Phi^{3/2}}{1+\frac12 \xi\Phi_{\xi}/\Phi}\displaystyle\frac{d\theta}{d\xi} \right)=0
\end{equation}
which for $n=3$ simplifies to
\begin{align}
 0=\frac{\theta^2\Big(2\alpha\theta(2+3\xi^2\theta^2)\theta'+6\alpha\xi\theta'^2-\xi\theta\Big)}{\xi}
 -\frac{2\theta'}{\xi}+(2\alpha\theta^3-1)\theta''
\end{align}
Let us recall that for $n=3$ the mass and radius of a white dwarf can be expressed as functions of the solutions and their first
zeros of the Lane-Emden equation:
\begin{align}
 \mathcal{M}&=-4\pi\left(\frac{K}{\pi G}\right)^{\frac{3}{2}}\xi_R^2\theta'(\xi_R),\\
 R&=\left(\frac{K}{\pi G}\right)^\frac{1}{2}\rho_c^{-\frac{1}{3}}\xi_R,
\end{align}
where $\xi_R$ is the first zero of $\theta(\xi)$. The function $\theta(\xi)$ is a solution of the Lane-Emden equation, $\rho_c$ the central density, $G$ the gravitational constant while 
$K=1.2435\times10^{15}\mu_e^{-4/3}$ cgs for $n=3$. Since the eventual differences in masses and in radii with respect to a theory of gravity are carried by 
$\xi_R$ and $\theta'(\xi_R)$, we may write down
\begin{align}
 M&=0.7212475 M_\odot \left(\frac{2}{\mu_e}\right)^2 \xi^2_R |\theta'(\xi_R)|,\\
 R&=0.48515\times10^4\left( \frac{10^6\text{g/cm}^{3} }{\rho_c} \right)^{1/3} \left(\frac{2}{\mu_e}\right)^{2/3}\xi_R \text{ km}
\end{align}
We will consider here the Chandrasekhar model, and thus the mean molecular weight per electron is $\mu_e=2$.

In the table \ref{tab} we have obtained the radii and masses for a few values of the parameter $\alpha=\kappa^2 c^2 \beta\rho_c$, were $\beta$ is the Starobinsky parameter. Let us notice 
that $\alpha$ depends on the central density $\rho_c$ which is a feature of the metric-affine (Palatini) gravity. Because of the choice of the Starobisnky 
model, the conformal function for the polytropic EoS is expressed as $\Phi=1-2\alpha\theta^3$ in the Lane-Emden variables and as discussed in \cite{aneta2}, the 
function is singular for $\alpha=1/2$. Therefore, the range of the values for the parameter is $\alpha\in(-\infty;0.5)$ because above the value $\alpha=0.5$ one
obtains non-physical values.

It should be also recalled that in \cite{aneta3} the stability of a non-relativistic polytrope was discussed for the model we have focused on. Thus, it was shown that for $n=3$ one deals with a stationary point if the parameter $\alpha$ is negative. This is also visible from the table: for positive values of $\alpha$ the radius' value increases while the mass decreases, decreasing density of the star.

Let us just comment the main difference between the Newtonian limit and Lane-Emden equation given for metric $f(R)$ gravity in \cite{cap1,cap2,cap3} - the equation is much complicated and it does not recover the well-known solutions in the GR limit which is however not a case in Palatini $f(\mathcal{R})$ gravity. Further works in this area, as it was presented for the metric $f(R)$ gravity in \cite{nn1,kal}, will allow to constrain Palatini gravity.

\begin{table}
\centering
\caption{Numerical values of the first zeros $\xi_R$ and the functions $\xi_R^2|\theta'(\xi_R)|$, together with the radii $R$ 
$\Big(\times10^4\left( \frac{10^6\text{g/cm}^{3} }{\rho_c} \right)^{1/3}\text{ km}\Big)$
and masses $M=\mathcal{M}/M_\odot$
for various values of $\alpha$.}
\label{tab}
\begin{tabular}{|c||c c|c c|}
\hline
$\alpha$ & $\xi_R$ & $\xi_R^2|\theta'(\xi_R)|$ & $R$  & $M$  \\
\hline\hline
0.49 & 10.32 & 1.396  & 5.008 & 1.0072   \\
0.4  & 9.523 & 1.407  &  4.62 & 1.01471   \\
0.3  & 8.486 & 1.468 &  4.12 & 1.05912   \\
0.2  & 7.651 & 1.595 &  3.712& 1.15068   \\
0.15  & 7.354 & 1.682  &  3.568 & 1.2128  \\
0.05 & 6.985 & 1.894  &  3.389 & 1.36579   \\
\hline
0    & 6.897 & 2.018 &  3.347 & 1.457   \\
\hline
-0.05   & 6.863 & 2.155 &  3.3 & 1.55413   \\
-0.1   & 6.875 & 2.30 &  3.34 & 1.66139   \\
-0.15  & 6.932 & 2.465 &  3.363 & 1.7778   \\
-0.2  & 7.027 & 2.64 &  3.41 & 1.90393   
\\-0.3  & 7.322 & 3.035 &  3.55 & 2.18932  
\\-0.4  & 7.748 & 3.505 &  3.759 & 2.52798 
\\-0.5  & 8.30119 & 4.07 &  4.03 & 2.93518     \\
\hline
\end{tabular}
\end{table}

\section{Conclusions}
The aim of this letter was to introduce a few aspects related to white dwarfs which
lie in a field of interest to the author. Thus, we have briefly discussed the age of such objects and super-Chandrasekhar mass. The first problem is strongly linked to the early evolution of white dwarf's progenitor star and cooling process of the star's compact remnant. It was already demonstrated in various works that both phenomena are strongly affected by modifications of the gravitational counterpart of the stellar equations. On the other hand, the second issue was broadly examined in modified gravity confirming that indeed super luminous SNIas could be caused by white dwarfs which exceed the (General Relativity) Chandrasekhar limit since the limit significantly differs in many theories of gravity.

Therefore studying those extraordinary objects in different gravitational frameworks may not only answer the posted questions but also shed light on the
dynamical effects introduced by modified gravity because of new effective interactions in the matter sector which may end up altering the
effective pressure generated by the matter sources. The high densities and pressures achievable by white
dwarfs turn them into excellent laboratories to confront the standard and modified theories with observations.

In the last part of this paper we obtained Chandrasekhar limit for quadratic Palatini $f(\mathcal R)$ gravity which has not been yet studied in this theory of gravity. We have shown that for the negative parameter $\alpha$ one deals with limited masses bigger than GR equivalent and thus, after more detailed investigations and realistic description, Palatini gravity could also contribute to explanation the mentioned problems related to white dwarfs and stars' modeling.

\section*{Acknowledgments}
  The work is supported by the European Union throughthe ERDF CoE grant TK133.


\end{document}